\documentclass[aps,prl,twocolumn,superscriptaddress,groupedaddress]{revtex4}  
\usepackage{graphicx}  
\usepackage{dcolumn}   
\usepackage{bm}        
\usepackage{amssymb}   

\begin{document}



\title{Ratio between two $\Lambda$ and $\bar{\Lambda}$ production 
mechanisms in $p$ scattering}
\author{B. Hoeneisen}
\affiliation{Universidad San Francisco de Quito, Quito, Ecuador}
\date{April 18, 2016}

\begin{abstract}
We consider $\Lambda$ and $\bar{\Lambda}$ production in a wide
range of proton scattering experiments. The produced
$\Lambda$ and $\bar{\Lambda}$ may or may not contain a diquark
remnant of the beam proton. The ratio of these two production
mechanisms is found to be a simple universal function
$r = [ \kappa/(y_p - y) ]^i$ of the rapidity difference $y_p - y$ of
the beam proton and the produced $\Lambda$ or $\bar{\Lambda}$,
valid over four orders of magnitude, from $r \approx 0.01$ to
$r \approx 100$, with 
$\kappa = 2.86 \pm 0.03 \pm 0.07$, and $i = 4.39 \pm 0.06 \pm 0.15$.
\end{abstract}

\maketitle


The $\bar{\Lambda}/\Lambda$ production ratio measured in a wide
range of proton scattering experiments $p Z \rightarrow \Lambda (\bar{\Lambda}) X$
has been found to be
a universal function $f(y_p - y)$ of ``rapidity loss" $y_p - y$,
where $y_p$ and $y$ are, respectively, the rapidities of the beam proton and
the produced $\Lambda$ or $\bar{\Lambda}$ \cite{LHCb, l}.
The function $f(y_p - y)$ is observed to be independent,
or depends only weakly, on the total center of mass energy
$\sqrt{s}$ of the two colliding hadrons in the range
0.024 to 7 TeV, on the target $Z = p, \bar{p}$, Be or Pb,
on the transverse momentum $p_T$ of the 
$\Lambda$ or $\bar{\Lambda}$, or on sample composition \cite{l}. 
We consider the picture in which an $s$ quark produced
in the scattering may coalesce with a $ud$ diquark remnant of the
beam proton and produce a $\Lambda$ \cite{l,x,x2,x3,x4}.

\begin{figure}
\includegraphics[scale=0.35]{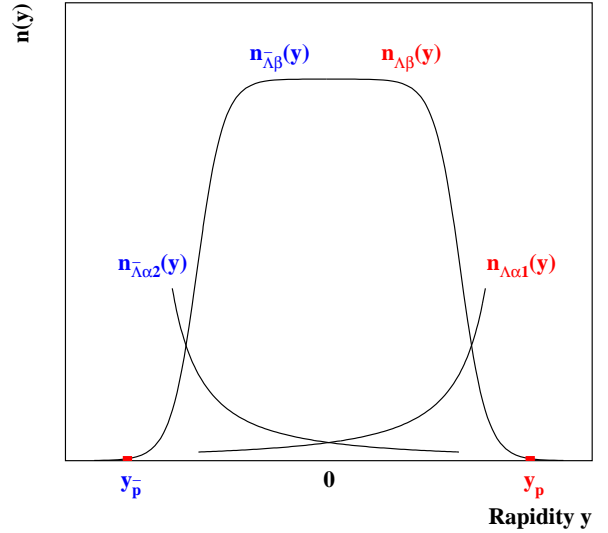}
\caption{\label{ny} Distributions of $\Lambda$ and $\bar{\Lambda}$
production as a function of rapidity $y$
in the center of mass frame of $p \bar{p}$ scattering.
The beam rapidities are respectively $y_p$ and $y_{\bar{p}}$.
The sub-index $\alpha$ ($\beta$) denotes $\Lambda$'s or $\bar{\Lambda}$'s
containing (not containing) a beam diquark remnant.
Lowering the total center of mass energy $\sqrt{s}$ translates
$y_p$ and $n_{\Lambda \alpha 1}(y)$ left, and translates
$y_{\bar{p}}$ and $n_{\bar{\Lambda} \alpha 2}(y)$ right.
}
\end{figure}

Let $n_\Lambda(y)$ and $n_{\bar{\Lambda}}(y)$ be the distributions 
of $\Lambda$'s and $\bar{\Lambda}$'s as a function
of rapidity $y$ in the center of mass frame of the two colliding hadrons.
Rapidity is defined so that the $p$ beam has positive rapidity $y_p$.
We write these distributions as follows:
\begin{eqnarray}
n_{\Lambda}(y) & = & n_{\Lambda \alpha 1}(y) + n_{\Lambda \alpha 2}(y) 
  + n_{\Lambda \beta}(y), \nonumber \\
n_{\bar{\Lambda}}(y) & = & n_{\bar{\Lambda} \alpha 1}(y) + n_{\bar{\Lambda} \alpha 2}(y) 
  + n_{\bar{\Lambda} \beta}(y), 
\label{n}
\end{eqnarray}
where $n_{\Lambda \alpha 1}(y)$ is the distribution of $\Lambda$'s containing
a diquark remnant of beam 1, 
$n_{\Lambda \alpha 2}(y)$ is the distribution of $\Lambda$'s containing
a diquark remnant of beam 2,
and $n_{\Lambda \beta}(y)$ is the distribution of $\Lambda$'s containing
no beam remnant, and similarly for $\bar{\Lambda}$'s.
Production mechanism $\beta$ has no memory of the beams and hence
$n_{\Lambda \beta}(y) = n_{\bar{\Lambda} \beta}(y) \equiv n_{\beta}(y) = n_{\beta}(-y)$.
The distribution $n_{\beta}(y)$ is approximately independent of $y$ within the
``rapidity plateau" $-y_\textrm{max} < y < y_\textrm{max}$  
as shown schematically in Fig. \ref{ny}. See also Fig. 2 of \cite{l}, and
Fig. 50.4 of \cite{PDG}.
We assume that beam 1 is a proton beam, so $n_{\bar{\Lambda} \alpha 1} = 0$.
If beam 2 is a $p$ beam, $n_{\bar{\Lambda} \alpha 2} = 0$.
If beam 2 is a $\bar{p}$ beam, $n_{\Lambda \alpha 2} = 0$.

We now consider events with $y > y_\textrm{min}$ so that
$n_{\bar{\Lambda} \alpha 2}$ and $n_{\Lambda \alpha 2}$
can be neglected. The $\bar{\Lambda}/\Lambda$ production ratio
can then be written as:
\begin{equation}
f = \frac{n_{\beta}(y)}{n_{\beta}(y) + n_{\Lambda \alpha 1}(y)} \equiv \frac{1}{1 + r},
\label{f}
\end{equation}
where the ratio, $r$, of the yields of the 
two production mechanisms $\alpha$ and $\beta$ is
\begin{equation}
r \equiv \frac{n_{\Lambda \alpha 1}(y)}{n_{\beta}(y)} = \frac{1}{f} - 1.
\label{r}
\end{equation}
The purpose of this note is to point out that 
the ratio $r$ can be fit
over four orders of magnitude, from $r \approx 0.01$ to
$r \approx 100$, with a simple universal function 
with only two parameters $\kappa$ and $i$:
\begin{equation}
r = \left[ \frac{\kappa}{y_p - y} \right]^i.
\label{i}
\end{equation}

\begin{figure}
\includegraphics[scale=0.35]{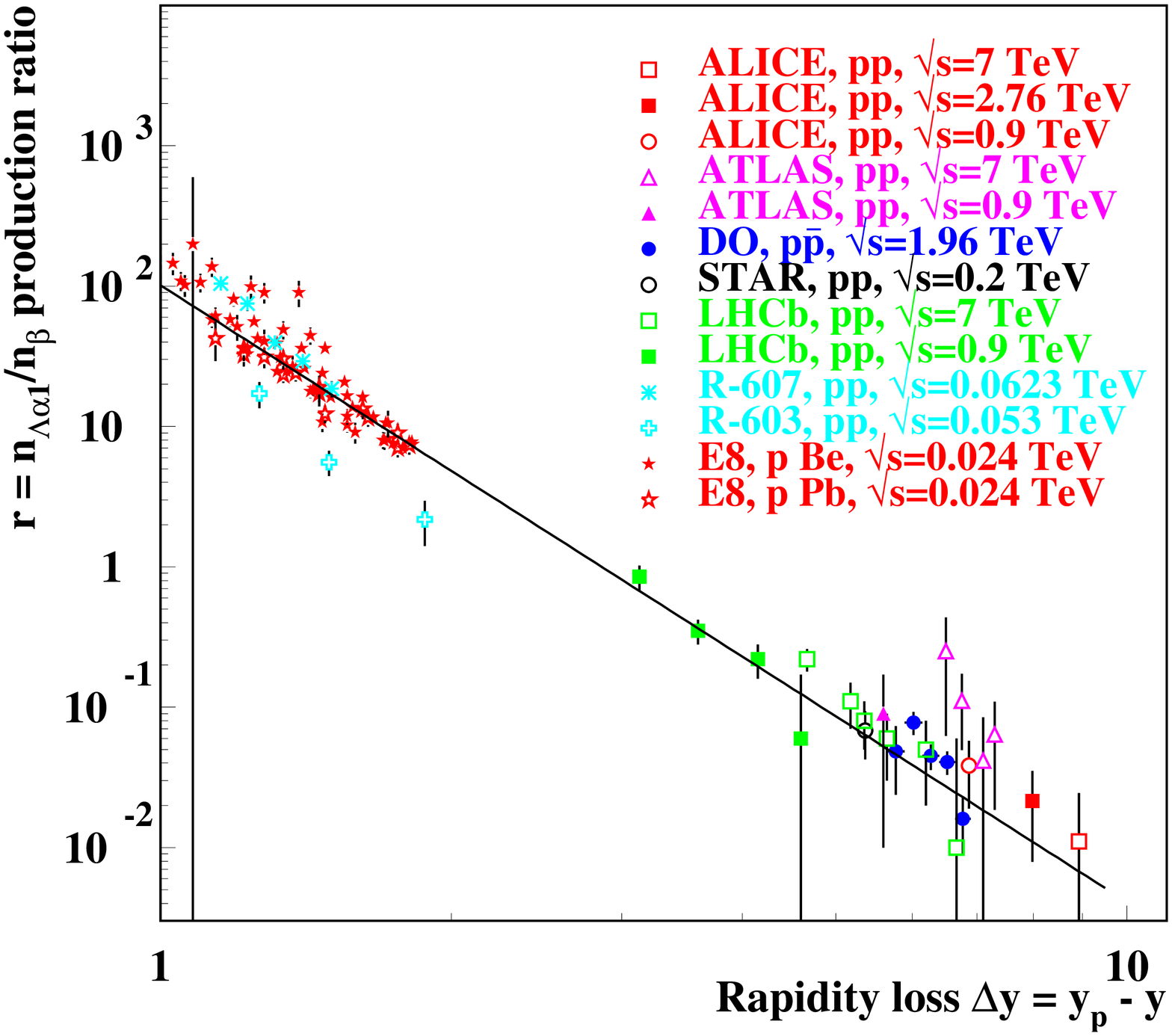}
\caption{\label{ri} Ratio 
$r \equiv n_{\Lambda \alpha 1}(y) / n_{\beta}(y)$
of $\Lambda$'s produced
with and without a beam proton remnant, as a function of
rapidity loss $y_p - y$, obtained by the experiments 
listed \cite{ALICE, ATLAS, l, STAR, ISR607, ISR, E8}. 
The fit shown has the form
$r = [ \kappa/(y_p - y) ]^i$ with 
$\kappa = 2.862$ and $i = 4.393$.}
\end{figure}

Figure \ref{ri} presents the ratios $r = 1/f - 1$ measured in a wide range of proton
scattering experiments. The data points with $y < 0.75$
of the D\O\ $p \bar{p}$ experiment were omitted because for them we
can not neglect $n_{\bar{\Lambda} \alpha 2}(y)$.
The data point of the STAR $pp$ experiment has $y = 0$, so 
$n_{\Lambda \alpha 1} = n_{\Lambda \alpha 2}$. Therefore we have divided
$1/f - 1$ by 2. 

The parameters $\kappa$ and $i$ have a simple interpretation and can be read off
the $\log{(y_p - y)}$  vs $\log{r}$ graph in
Fig. \ref{ri}: $\kappa \approx 2.8$ is the rapidity loss at which $r = 1$,
and $-i \approx -4.4$ is the slope of the straight line in Fig. \ref{ri}. 
For $y_p - y < \kappa$ production mechanism
$\alpha$ dominates. For $y_p - y > \kappa$ production mechanism
$\beta$ dominates.

The fit to all of the data in Fig. \ref{ri} obtains
$\kappa = 2.79 \pm 0.03$ and $i = 4.54 \pm 0.08$ with $\chi^2 = 637$ for 121 degrees of freedom. 
The large $\chi^2$ is due to  
tension between the data points of different experiments
as can be seen in Fig. \ref{ri}.
Omitting the R-603 and R-607 measurements, 
which have some data points off the rapidity plateau,
obtains
$\kappa = 2.86 \pm 0.03$ and $i = 4.39 \pm 0.06$ with $\chi^2 = 342$ for 102 degrees of freedom.
This is the fit shown in Fig. \ref{ri}.
Fitting only the E8 Pb data points where production mechanism 
$\alpha$ dominates obtains
$\kappa = 2.93 \pm 0.15$ and $i = 4.06 \pm 0.30$ with $\chi^2 = 10.6$ for 13 degrees of freedom.
Fitting all data with $y_p - y > 2.8$,
where production mechanism $\beta$ dominates, obtains
$\kappa = 2.94 \pm 0.10$ and $i = 4.23 \pm 0.25$ with $\chi^2 = 37$ for 32 degrees of freedom.
In conclusion, we see no significant departure from Eq. (\ref{i}) at either end of the data range.
Our final estimate from several fits is
\begin{equation}
\kappa = 2.86 \pm 0.03 \pm 0.07,
i = 4.39 \pm 0.06 \pm 0.15,
\end{equation}
where the first uncertainty is statistical from the fit, and
the second uncertainty is systematic and accounts for 
different data selections for the fits.

From $\sqrt{s} = 0.024$ to 7 TeV the cross-sections
$\sigma(pp)$ and $\sigma(p \bar{p})$, and the width
of the rapidity plateau $2 y_\textrm{max}$, increase
by approximately a factor 2 \cite{PDG}, so $n_\beta(y)$ is
approximately independent of $\sqrt{s}$ and $y$
on the rapidity plateau.
We conclude that the probability density that a
$p$ scatters and becomes a $\Lambda$ with rapidity $y$
is proportional to $[ \kappa/(y_p - y) ]^i$.
This result should also be valid for $\Lambda_c$,
$\Lambda_b$, $\Sigma^+$, etc.

I thank my colleagues in the D0 Collaboration for their 
comments and inspiration to undertake this analysis.

\end{document}